\documentclass[amsmath,amssymb,prc,final,showpacs,twocolumn]{revtex4}
\usepackage{bm,hyphenat,xspace}
\usepackage{graphicx,epsfig}

\newcommand{\Eq}{Eq.}
\newcommand{\Eqs}{Eqs.}
\newcommand{\Fig}{Fig.}

\newcommand{\Ref}{Ref.}
\newcommand{\Refs}{Refs.}

\newcommand {\mbf}[1]{{\mathbf{#1}}}

\newcommand {\mcu}{\mathcal{U}}
\newcommand {\mcv}{\mathcal{V}}
\newcommand {\mcg}{\mathcal{G}}
\newcommand {\mct}{\mathcal{T}}

\newcommand{\He}{{}^3\mathrm{He}}
\newcommand{\Hh}{{}^3\mathrm{H}}
\newcommand{\nH}{n\text{-}{}^3\mathrm{H}}
\newcommand{\pHe}{p\text{-}{}^3\mathrm{He}}
\newcommand{\pH}{p\text{-}{}^3\mathrm{H}}
\newcommand{\nHe}{n\text{-}{}^3\mathrm{He}}
\newcommand{\pd}{p\text{-}d}
\newcommand{\nd}{n\text{-}d}
\newcommand{\dd}{d\text{-}d}

\begin{document}

\title {Four-nucleon scattering: 
\textit{ab initio} calculations in momentum space}
 
\author{A.~Deltuva} 
\email{deltuva@cii.fc.ul.pt}
\affiliation{Centro de F\'{\i}sica Nuclear da Universidade de Lisboa, 
P-1649-003 Lisboa, Portugal }

\author{A.~C.~Fonseca} 
\affiliation{Centro de F\'{\i}sica Nuclear da Universidade de Lisboa, 
P-1649-003 Lisboa, Portugal }

\received{November 9, 2006}

\begin{abstract}
The four-body equations of Alt, Grassberger and Sandhas are solved
for $\nH$ scattering at energies below
three-body breakup threshold using various realistic interactions
including one derived from chiral perturbation theory.
 After partial wave decomposition the
equations are three-variable integral equations that are solved
numerically without any approximations beyond the usual
discretization of continuum variables on a finite momentum mesh.
Large number of two-, three- and four-nucleon partial waves are considered 
until the convergence of the results is obtained. The total
$\nH$ cross section data in the resonance region is not
described by the calculations which confirms previous findings by
other groups. Nevertheless the numbers we get are slightly higher
and closer to the data than previously found and depend on the choice
of the  two-nucleon potential.
Correlations between the $A_y$ deficiency in $\nd$ elastic
scattering and the total $\nH$ cross section are studied. 

\end{abstract}

 \maketitle

%%%%%%%%%%%%%%%%%%%%%%%%%%%%%%%%%%%%%%%%%%%%%%%%%%%%%%%%%%%%%%%%%%%%%%%%%%%%%%%
\section{Introduction \label{sec:intro}}

The four-nucleon $(4N)$ scattering problem gives rise to the simplest set
of nuclear reactions that shows the complexity of heavier systems. 
The neutron-$\Hh$ $(\nH)$ and proton-$\He$ $(\pHe)$ scattering is 
dominated by the total isospin $\mct = 1$ states while deuteron-deuteron 
$(\dd)$ scattering by the $\mct = 0$ states; the reactions $\nHe$ and
$\pH$ involve both $\mct = 0$ and $\mct = 1$ and are coupled to
$\dd$ in $\mct = 0$. Due to the charge dependence of the hadronic and 
electromagnetic interaction a small admixture of $\mct=2$ states 
is also present.
In $4N$ scattering the Coulomb interaction is paramount not only
to treat $\pHe$ but also to separate the $\nHe$ threshold from
$\pH$ and at the same time avoid a second excited state of the $\alpha$
particle a few keV bellow the lowest scattering threshold.
All these complex features make the $4N$ scattering problem not only a
natural theoretical laboratory to test different force models of the
nuclear interaction, but also the next step in the pursuit of very accurate
{\em ab initio} calculations of the $N$-body scattering problem after
the extensive work on the three-nucleon $(3N)$ system that has taken
place in the past twenty years by several groups
\cite{gloeckle:96a,golak:05a,kievsky:01a}.

In \Refs~\cite{fonseca:99a,fonseca:02a} all the reactions mentioned above
were studied in the framework of Alt, Grassberger, and Sandhas (AGS) 
equations \cite{grassberger:67} using the
rank one representation of realistic  two-nucleon $(2N)$
force models together with a high rank representation of all 
$3N$ subsystem amplitudes; the Coulomb interaction 
was neglected. This led to one-variable integral equations whose predictive
power was limited to the quality of the involved approximations. The
calculations showed large discrepancies with data, namely nucleon 
analyzing power $A_y$ in $\nHe$ scattering, tensor observables in  
$^2\mathrm{H}(\vec{d},n)\He$ and
$^2\mathrm{H}(\vec{d},d)^2\mathrm{H}$ and the differential cross section for
$^2\mathrm{H}(d,n)\He$, but one surprising success in describing
the total cross section $\sigma_t$ for $\nH$ scattering in the
resonance region where at neutron lab energy $E_n \simeq 3.5 $ MeV $\sigma_t$
rises to about 2.45 b \cite{phillips:80}. Calculations by the Grenoble group
\cite{cieselski:99} using coordinate-space solutions of the Faddeev-Yakubovsky
equations \cite{yakubovsky:67} showed, on the contrary, that realistic 
interactions missed the total cross section peak by at least 0.2 b. Although
these calculations carried out no approximation on the treatment of
the $2N$ interaction, they were limited vis-\'{a}-vis 
\Ref~\cite{fonseca:99a} on the number of $3N$ and $4N$ partial waves.  

Although the issue was recently clarified \cite{lazauskas:04a,lazauskas:05a} by
comparing to an independent calculation by the Pisa group that uses the
Kohn variational method, together with hyperspherical harmonics, further
studies based on the AGS equations are needed to settle this important problem
because some of the results by the Grenoble and Pisa groups may be still of 
limited accuracy given the number of included $2N$, $3N$, and $4N$ partial 
waves. Further investigations are also needed for the understanding of other 
$4N$ reactions such as  $\pHe$, $\nHe$ and $\dd$ where large discrepancies 
with data were previously found. 
One fundamental  issue underlying four-nucleon physics is the existence
of correlations between $3N$ and $4N$ observables. One of
the best known is the Tjon line \cite{tjon:75} which correlates the binding
energies of $\Hh$ with ${}^4\mathrm{He}$; another one involves the
triton binding energy and the singlet (triplet) $\nH$ scattering length 
\cite{viviani:98a}. Nevertheless, other correlations may exist: one
could ask if the persistent $A_y$ problem in $\nd$ scattering is in any
way related to the failure to reproduce $\sigma_t$ in $\nH$
scattering in the resonance region, or to the $A_y$ problem in
$\pHe$ \cite{viviani:01a}; does resolving the former also solves the latter?

Therefore we present here a new numerical approach to the solution
of the AGS equations that is both numerically exact and extremely
fast in terms of CPU-time demand. Since the $2N$ transition matrix
(t-matrix) is treated exactly, the equations we solve are, after partial wave
decomposition, three-variable integral equations. The three Jacobi momentum
variables in $1+3$ and $2+2$ configurations are discretized on a finite mesh
and the number of $2N$, $3N$ and $4N$ partial waves increased up to what is 
needed for the full convergence of the observables. The 
present approach also allows for the inclusion of charge-dependent
interactions as well as $\Delta$ degrees of freedom that lead to an effective
$3N$ force. Furthermore, using the method recently proposed to treat the 
Coulomb force in $\pd$ elastic scattering and breakup 
\cite{deltuva:05a,deltuva:05c,deltuva:05d}, we have already obtained 
preliminary results for $\pHe$ elastic scattering observables
\cite{fonseca:fb18} with the Coulomb potential between the three protons 
included.

In Sec.~\ref{sec:eq} we discuss the integral equations we solve for
$\nH$  scattering and in Sec.~\ref{sec:res} we show
the results of our most complete calculations, leaving tests of benchmark to
the Appendix. Finally conclusions come in Sec.~\ref{sec:conc}.

\section{Equations \label{sec:eq}}

As initially proposed by Alt, Grassberger and Sandhas \cite{grassberger:67} 
and later reviewed for the
purpose of practical applications in \Ref~\cite{fonseca:87}, the four-particle
scattering equations may be written in a matrix form 
\begin{subequations} \label{eq:AGSmtr}
\begin{align}
\mcu = {} & \mcv + \mcv \, \mcg_0 \, \mcu, \\
\mcu |\Phi_{\rho_0} \rangle = {}& \mcv |\Psi_{\rho_0} \rangle, \\
|\Psi_{\rho_0} \rangle = {}& |\Phi_{\rho_0} \rangle + 
\mcg_0 \, \mcv |\Psi_{\rho_0}  \rangle ,
\end{align}
\end{subequations}
where 
$|\Phi_{\rho_0} \rangle$ is the initial channel state,
$|\Psi_{\rho_0} \rangle$ the full scattering state,
and $\rho_0$ defines the two-body entrance channel.
Both of them have 18 components, and the transition operator
$\mcu$ as well as $\mcv$ and $\mcg_0$ are $18 \times 18$ matrix operators
with components 
\begin{subequations} \label{eq:AGSdef}
\begin{align}
[\mcv]^{\sigma \rho}_{ij} = {} & (G_0\, t_i\, G_0)^{-1}\, 
\bar{\delta}_{\sigma \rho} \, \delta_{ij}, \\
[\mcg_0]^{\sigma \rho}_{ij} = {} & 
G_0\, t_i\, G_0\, U^{\sigma}_{ij} G_0\, t_j\, G_0\, \delta_{\sigma \rho}. 
\end{align}
\end{subequations}
As usual, $\sigma(\rho)$ denotes two-cluster partitions of $1+3$ or 
$2+2$ type and $i(j)$ the pair interactions.
$G_0$ is the four free particle Green's function, $t_i$ is the two-particle 
t-matrix embedded in four-particle space, 
$\bar{\delta}_{\sigma \rho} = 1 - \delta_{\sigma \rho}$, 
and $U^{\sigma}_{ij}$ are the subsystem transition operators 
\begin{gather}
U^{\sigma}_{ij} = G^{-1}_0 \, \bar{\delta}_{ij} + \sum_k \bar{\delta}_{ki}
\, t_k \, G_0 \, U^{\sigma}_{kj}
\end{gather}
of $1+3$ or $2+2$ type, depending on $\sigma$.  
If $\sigma$ is a $1+3$ partition, $U^{\sigma}_{ij}$ corresponds to the 
usual AGS transition matrix for the three interacting particles %$(i,j,k)$ 
that are internal to $\sigma$.
For $\sigma$ of $2+2$ type $U^{\sigma}_{ij}$ does not
correspond to any physical process.
The components of the initial/final two-cluster states 
$[|\Phi_{\rho_0}  \rangle ]^\rho_i = 
|\phi^{\rho_0}_i  \rangle  \delta_{\rho \rho_0}$
are the Faddeev components of the cluster bound state wave function 
times a plane wave of momentum $\mbf{p}_{\rho_0}$ between clusters whose
dependence is suppressed in our notation,
\begin{gather}
|\phi^{\rho_0}_i  \rangle = G_0 \, t_i 
\sum_k  \bar{\delta}_{ki} |\phi^{\rho_0}_k  \rangle.
\end{gather}

The great advantage of AGS equations over the Yakubovsky equations is that
on-shell matrix elements of $\mcu$ between initial $|\Phi_{\rho_0} \rangle $
and final $|\Phi_{\sigma_0} \rangle $ states with relative two-cluster
momenta $\mbf{p}_{\rho_0}$ and $\mbf{p}'_{\sigma_0}$
lead automatically to the corresponding scattering amplitudes 
\begin{subequations} \label{eq:Ton}
\begin{align}
\langle \mbf{p}'_{\sigma_0}| T^{\sigma_0\rho_0} |\mbf{p}_{\rho_0} \rangle
  = {} & \langle \Phi_{\sigma_0}| \mcu |\Phi_{\rho_0}  \rangle   \\
  = {} & \sum_{ij} \langle  \phi^{\sigma_0}_j | \mcu^{\sigma_0 \rho_0}_{ji} 
| \phi^{\rho_0}_i \rangle .
\end{align}
\end{subequations}

For four identical particles the AGS equations reduce to $2 \times 2$
matrix equations since there are only two distinct partitions, one of 
$1+3$ type and one of $2+2$ type, which we choose to be
(12,3)4 and (12)(34); in the following we denote them by 
$\alpha =1$ and $\alpha=2$, respectively. 
In this case the equations may be conveniently written using the 
permutation operators $P_{ab}$ of particles $a$ and $b$ as it was done first 
in \Refs~\cite{kamada:92a,gloeckle:93a} for the four-nucleon bound state.
After the symmetrization of the four-nucleon scattering equations 
\eqref{eq:AGSmtr} we obtain equations of the same form
but with new definitions for the symmetrized operators
\begin{subequations} \label{eq:AGSsymm}
\begin{align}
\mcv^{\alpha \beta} = {} & (G_0\, t \, G_0)^{-1}\, 
(\bar{\delta}_{\alpha \beta} - \delta_{\beta 1} P_{34}) , \\
\mcg_0^{\alpha \beta} = {} & 
G_0\, t\, G_0\, U^{\alpha} G_0\, t\, G_0\, \delta_{\alpha \beta}.
\end{align}
\end{subequations}
Here $t$ is the pair (12) t-matrix, $U^\alpha$
the symmetrized 1+3 or 2+2 subsystem transition operators 
\begin{gather} \label{eq:AGSsub}
U^\alpha =  P_\alpha G_0^{-1} + P_\alpha t\, G_0 \, U^\alpha,
\end{gather}
and $P_\alpha$ the permutation operators given by
\begin{subequations} \label{eq:P}
\begin{align}
P_1 = {} & P  =  P_{12}\, P_{23} + P_{13}\, P_{23},\\
P_2 = {} & \tilde{P}  =  P_{13}\, P_{24}.
\end{align}
\end{subequations}
The basis states are antisymmetric under exchange of two particles in 
subsystem (12) for $1+3$ partition and in (12) and (34) for $2+2$ partition.
The symmetrized initial/final two-cluster state components are
\begin{gather} \label{eq:phi}
|\phi^{\beta} \rangle = G_0 \, t  P_{\beta} |\phi^{\beta} \rangle.
\end{gather}
The scattering amplitudes are obtained as
\begin{gather}
\langle \mbf{p}'_{\alpha}| T^{\alpha\beta} |\mbf{p}_{\beta} \rangle
  = S_{\alpha\beta} 
\langle  \phi^{\alpha} | \mcu^{\alpha\beta}| \phi^{\beta} \rangle,
\end{gather}
where $S_{\alpha\beta}$ is a symmetrization factor;
$S_{\alpha\alpha} = S_{\alpha}$  is the number of pairs 
internal to the partition $\alpha$, i.e., $S_1 = 3$ and $S_2 = 2$,
and $S_{12} = 2 S_{21} = 2\sqrt{3}$.

Since the present paper is confined to $\nH$ scattering, we write down
explicitly only the equations for the
$1+3 \to 1+3$ and $1+3 \to 2+2$ transition operators
\begin{subequations}  \label{eq:AGS}   
\begin{align}  
\mcu^{11}  = {}&  -(G_0 \, t \, G_0)^{-1}  P_{34} -
P_{34} U^1 G_0 \, t \, G_0 \, \mcu^{11}  \nonumber \\ 
{}& + U^2   G_0 \, t \, G_0 \, \mcu^{21}, \label{eq:U11}  \\
\label{eq:U21}
\mcu^{21} = {}&  (G_0 \, t \, G_0)^{-1}  (1 - P_{34})
+ (1 - P_{34}) U^1 G_0 \, t \, G_0 \, \mcu^{11}.
\end{align}
\end{subequations}
The equations  coupling $\mcu^{12}$ and $\mcu^{22}$ share an identical 
kernel but  different inhomogeneous terms.

\begin{figure}
\begin{center}
\includegraphics[scale=0.25]{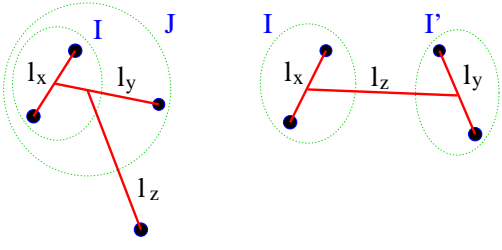}
\end{center} 
\caption{ \label{fig:1} (Color online)
Angular momentum quantum numbers for $1+3$ and $2+2$ basis states.}
\end{figure}

After the partial wave expansion \Eqs~\eqref{eq:AGS} form a set of 
coupled integral equations with three variables corresponding to the
Jacobi momenta $k_x , \, k_y$ and $k_z$; the associated orbital angular 
momenta are denoted by $l_x$, $l_y$, and $l_z$, respectively.
They are  depicted in \Fig~\ref{fig:1} for $1+3$ and $2+2$ configurations 
together with the pair total angular momentum $I$ and the 
the three-particle subsystem total angular momentum $J$.
The states of total angular momentum  $\mathcal{J}$ are defined as  
$ | k_x \, k_y \, k_z   
[l_z ( \{l_y [(l_x S_x)I \, s_y]S_y \} J s_z ) S_z] \,\mathcal{J M} \rangle $ 
for the $1+3$ configuration and 
$|k_x \, k_y \, k_z  (l_z  \{ (l_x S_x)I\, [l_y (s_y s_z)S_y] I' \} S_z)
\mathcal{ J M} \rangle $ for the $2+2$,  where $s_y$ and $s_z$ are the spins
of nucleons 3 and 4, and $S_x$, $S_y$, and $S_z$ are channel spins
of two-, three-, and four-particle system.
In all calculations $I$ 
and $I'$ run over the same set of quantum numbers.

By the discretization of the momentum variables the integral equations
may be turned into a system of linear equations but the direct solution
is not possible because of the huge dimension. 
Therefore, in close analogy with three-nucleon scattering,
we calculate the Neumann series for the on-shell matrix elements
of the transition operators \eqref{eq:AGS} and sum by the Pad\'{e} method
\cite{baker:75a}. The Pad\'{e} summation
algorithm we use is described in \Ref~\cite{chmielewski:03a}.
We work with the half-shell transition operators in the form 
\begin{gather}
|X_{\alpha\beta} \rangle = G_0\, \mcu^{\alpha\beta} |\phi^{\beta} \rangle
\end{gather}
such that the on-shell elements are
$\langle  \phi^{\alpha} | \mcu^{\alpha\beta}| \phi^{\beta} \rangle =
\langle \xi_{\alpha} | X_{\alpha\beta} \rangle $ with the auxiliary states
$|\xi_{\alpha} \rangle = G_0^{-1} | \phi^{\alpha} \rangle 
= t  P_{\alpha} |\phi^{\alpha} \rangle $. 
Defining $Q_\alpha = G_0 \, U^\alpha G_0 \, t $ and
using \Eq~\eqref{eq:phi} for the inhomogeneous terms in order to eliminate
$(G_0 \, t \, G_0)^{-1} $, \Eqs~\eqref{eq:AGS} become
\begin{subequations}  \label{eq:AGS-X}   
\begin{align}  
|X_{11}\rangle  = {}&  - P_{34} P_1 G_0| \xi_{1} \rangle -
P_{34} Q_1 |X_{11}\rangle
+  Q_2  |X_{21}\rangle, \label{eq:X11}  \\
\label{eq:X21}
|X_{21}\rangle = {}&  (1 - P_{34}) P_1 G_0 | \xi_{1} \rangle
+ (1 - P_{34}) Q_1 |X_{11}\rangle.
\end{align}
\end{subequations}
In practical calculations, in order to accelerate the convergence of the
Pad\'{e} summation, it is advantageous to substitute 
\Eq~\eqref{eq:X21} into \Eq~\eqref{eq:X11} yielding the Neumann series
\begin{subequations}  \label{eq:NeumannX}   
\begin{align}  
|X_{\alpha\beta} \rangle = {}&  \sum_{n=0}^{\infty} 
|X_{\alpha\beta}^{(n)} \rangle, \\
|X_{21}^{(0)} \rangle = {}& (1 - P_{34}) P_1 G_0| \xi_{1} \rangle , \\
|X_{11}^{(0)} \rangle = {}& -P_{34} P_1 G_0| \xi_{1} \rangle 
+ Q_2  |X_{21}^{(0)}  \rangle, \\
|X_{21}^{(n)} \rangle = {}& (1 - P_{34}) Q_1 |X_{11}^{(n-1)} \rangle, \\
|X_{11}^{(n)} \rangle = {}& -P_{34} Q_1 |X_{11}^{(n-1)} \rangle
+ Q_2  |X_{21}^{(n)}  \rangle,
\end{align}
\end{subequations}
which requires $1+3$ and $2+2$ subsystem transition operators $U^{\alpha}$,
contained in  $Q_{\alpha}$, fully off-shell at different energies. 
Explicit calculation of $U^{\alpha}$ is not only very time consuming but also
requires large storage devices. Therefore,
except at the bound state poles, we do not calculate the 
full off-shell transition matrices  $U^{\alpha}$ explicitly.
Instead, we rewrite \Eq~\eqref{eq:AGSsub} as a Neumann series
\begin{gather} \label{eq:Neumann}
U^\alpha =  \sum_{r=0}^{\infty} (P_{\alpha} t\, G_0)^r\, P_{\alpha} G_0^{-1}
\end{gather}
resulting a corresponding Neumann series for the solution vectors
in \Eqs~\eqref{eq:NeumannX}, i.e.,
\begin{subequations}  \label{eq:NeumannQ}   
\begin{align}  
Q_\alpha  |X_{\alpha\beta}^{(n)} \rangle = {} &
\sum_{r=1}^{\infty} |X_{\alpha\beta}^{(n,r)}\rangle, \\
|X_{\alpha\beta}^{(n,0)}\rangle = {} & |X_{\alpha\beta}^{(n)}\rangle, \\
|X_{\alpha\beta}^{(n,r)}\rangle = {} &  P_{\alpha} G_0 \, t \,
|X_{\alpha\beta}^{(n,r-1)}\rangle, \label{eq:NeumannQ3}
\end{align}
\end{subequations}
where the summation again has to be performed using the Pad\'{e} method.
Usually, 6 to 18 Pad\'{e} iteration steps are required for the
convergence in \Eqs~\eqref{eq:NeumannX} - \eqref{eq:NeumannQ}.
At the bound state poles the subsystem transition operators are
\begin{gather} \label{eq:Bpole}
U^\alpha =  P_\alpha |\xi_\alpha \rangle 
\frac{S_\alpha}{E + i0 - h_0^z - E_{B}} \langle \xi_\alpha | P_\alpha,
\end{gather}
where $E$ is the available four-nucleon energy, $E_B$ the binding energy,
and $h_0^z$ the kinetic energy operator for the relative motion of
the two clusters.

Thus, compared to the calculation of full off-shell $U^{\alpha}$,
the method we are using avoids storage problems and also significantly
reduces the number of required floating point operations, since 
it is essentially a calculation of \emph{half-shell} matrix elements
for a number of driving terms that are considerably fewer than
the linear dimension of the discretized $U^{\alpha}$.
A further advantage is that the matrices corresponding to the operators 
$P_{\alpha}$,  $G_0$  and $t$ in \Eq~\eqref{eq:NeumannQ3} have 
block-diagonal structure whereas  $U^{\alpha}$ is a full matrix.

The calculation of the Neumann series \eqref{eq:NeumannQ} for $\alpha=1$
is what we are doing in three-nucleon scattering 
and is described in great detail in \Refs~\cite{deltuva:03a,deltuva:phd}. 
The specific representation of the permutation
operator $P_1$ where the initial and final state momenta $k_y$ are chosen 
as independent variables requires the interpolation in the momentum
$k_x$ for the quantities on both sides of $P_1 G_0$, i.e., for $t$
or $\langle \xi_\alpha |$. Two interpolation methods using Chebyshev
polynomials and spline functions were used in \Ref~\cite{deltuva:phd};
in the context of four-nucleon equations where one has to work with
$1+3$ and $2+2$ basis states the spline interpolation is more convenient.

The calculation of the Neumann series \eqref{eq:NeumannQ} for $\alpha=2$
is straightforward because of the very simple form of the permutation
operator $P_2$.

Finally, the application of the permutation operator $P_{34}$ as well
as the transformation of $|X_{\alpha\beta}^{(n)} \rangle $ from $1+3$ basis
to $2+2$ or vice versa has a structure similar to that of $P_1$,
resulting in a similar treatment. 
The specific representation of $P_{34}$, i.e.,
\begin{gather}  \label{eq:P4}   
\begin{split}  
\langle k_x k_y k_z | P_{34} | k'_x k'_y k'_z \rangle
= {} & \frac{\delta (k_x - k'_x)} {k_x^2}
\int_{-1}^{1} dy \; P_{34}(k_z,k'_z,y) 
\\ & \times
\frac{\delta (k_y - \bar{k}_y(k_z,k'_z,y))} {k_y^2} 
\\ & \times
\frac{ \delta (k'_y - \bar{k}'_y(k_z,k'_z,y))} {{k'_y}^2},
\end{split}
\end{gather}
where the initial and final state momenta $k_z$ are chosen 
as independent variables, requires the interpolation in the momentum
$k_y$ for the quantities on both sides of $P_{34}$
that are calculated on the mesh $\{ k_{i_y} \}$.
The dependence on the discrete quantum numbers is suppressed since
it is irrelevant for the consideration as well as the explicit
form of function $P_{34}(k_z,k'_z,y)$. $\bar{k}'_y(k_z,k'_z,y)$ and
and $\bar{k}_y(k_z,k'_z,y)$ are the initial and final state Jacobi momenta
$k_y$ expressed via $k'_z$, $k_z$, and the angle between them
$y = \hat{\mbf{k}}'_z \cdot \hat{\mbf{k}}_z$.
We use the spline interpolation again with the spline functions
$S_{i_y}(k)$ \cite{boor:78a,gloeckle:82a,press:89a}
such that for the function $f(k)$, given on the mesh $\{ k_{i_y} \}$,
the values at any $k$ may be obtained by
\begin{gather}
f(k) \approx \sum_{i_y} f(k_{i_y}) S_{i_y}(k).
\end{gather}
For $P_{34}$ acting on the vector $| Y \rangle $
we obtain the following result (as a distribution) 
\begin{gather}  \label{eq:P4X}
\begin{split}
\langle k_x k_y k_z | P_{34} | Y \rangle
= {} & \sum_{i_y} \frac{\delta (k_y - k_{i_y})}{k_y^2}
\int_0^{\infty} {k'_z}^2 d k'_z \int_{-1}^{1} dy \\ 
& \times S_{i_y}(\bar{k}_y(k_z,k'_z,y)) \, P_{34}(k_z,k'_z,y) \\ 
& \times \sum_{j_y} S_{j_y} (\bar{k}'_y(k_z,k'_z,y)) \,
\langle k_x  k_{j_y} k'_z | Y \rangle  \\
= {} & \sum_{i_y} \frac{\delta (k_y - k_{i_y})}{k_y^2} 
\tilde{Y}_{i_y}(k_x,k_z), 
\end{split}
\end{gather}
such that in the next step of the calculation,
where the $\langle k_x k_y k_z | P_{34} | Y \rangle$ has to be
multiplied by a smooth function $f(k_y)$ and integrated over $k_y$,
the result simply is the sum over the meshpoints $\{ k_{i_y} \}$ 
for the involved quantities,
\begin{gather}  \label{eq:ZP4X}
\int_0^{\infty} k_y^2 d k_y \, f(k_y)
\langle k_x k_y k_z | P_{34} | Y \rangle
= \sum_{i_y} f(k_{i_y}) \, \tilde{Y}_{i_y}(k_x,k_z).
\end{gather}
The integrations in \Eq~\eqref{eq:P4X} are performed using Gaussian
integration rules \cite{press:89a}. The bound state pole \eqref{eq:Bpole} 
is treated by the subtraction technique much like the deuteron pole
in the $\nd$ scattering \cite{deltuva:phd}.
Note that the representation of the operators $P_1$ and $P_{34}$
is different from the one used in \Refs~\cite{kamada:92a,nogga:02a} where
final state momenta $k_y$ and $k_z$ were chosen as independent variables.

\section{Results \label{sec:res}}

In order to calibrate our work we start by reproducing results of previous
calculations, in particular the binding energy of $^4\rm{He}$ obtained with
different  realistic $2N$ interactions by different groups
\cite{nogga:02a,kamada:01a,lazauskas:04a,viviani:06a} as well as the 
$\nH$ phase shifts obtained with Mafliet-Tjon potential by the Grenoble
group \cite{lazauskas:phd}.  Furthermore, we check the numerical stability
of our calculations.
These results are presented in the Appendix and show that
the present algorithm is numerically highly reliable and capable of 
reproducing previous published results. 

Next we study the convergence of our calculations in terms
of number of $2N$, $3N$, and $4N$ partial waves using the AV18 potential
\cite{wiringa:95a}  for the $2N$ interaction.
In the calculations presented here for the $\nH$ scattering we include
only the total isospin $\mct=1$ states, but, within $\mct=1$, we take into
account all couplings resulting from the charge dependence of the interaction.
Including $\mct=2$ states would yield an effect that is of 2nd order
in the charge dependence and, therefore, is expected to be extremely small
much like the effect of the total $3N$ isospin $T = \frac32$ states
in elastic $\nd$ scattering. Coupling to $\mct=2$ states is neglected
also in all previous calculations, but in configuration-space treatments
the isospin averaging within $\mct=1$ states is performed for the potential,
whereas we perform it for the t-matrix.

\begin{table*}[htbp] 
\begin{ruledtabular}
\begin{tabular}{l*{8}{c}}
& $0^+ \; (^1S_0)$  & $0^- \; (^3P_0)$ & $1^+ \; (^3S_1) $ &
  $1^- \; (^3P_1)$  & $1^- \; (^1P_1)$ & $1^- \; (\epsilon)$ & 
  $2^- \; (^3P_2)$  & $\sigma_t $
\\  \hline
$I \leq 1^+$
& -69.54 & 20.97 & -62.31 & 46.47 & 26.46 & -37.74 & 36.19 & 2.106 \\  
$I \leq 1 + {}^3P_2$ 
& -70.60 & 22.82 & -62.70 & 40.30 & 21.25 & -44.00 & 43.53 & 2.151 \\  
$I \leq 2 + {}^3D_3$ 
& -70.02 & 24.43 & -62.05 & 43.62 & 22.65 & -44.46 & 47.06 & 2.301 \\  
$I \leq 3 + {}^3F_4$ 
& -69.68 & 23.52 & -61.74 & 43.37 & 22.35 & -44.71 & 46.71 & 2.277 \\ 
$I \leq 4 + {}^3G_5$ 
& -69.63 & 23.62 & -61.69 & 43.54 & 22.38 & -44.69 & 47.03 & 2.288 \\ 
$I \leq 5 + {}^3H_6$ 
& -69.61 & 23.56 & -61.68 & 43.53 & 22.37 & -44.73 & 47.00 & 2.286
\end{tabular}
\end{ruledtabular}
\caption{ \label{tab:1} 
$\nH$ phase shifts, $1^-$ mixing parameter $\epsilon$ (in degrees)
and the total cross section $\sigma_t$ (in barns)
at $E_n = 4$ MeV neutron lab energy for increasing number 
of $2N$  partial waves and fixed $l_y, l_z \leq 4$, $J \leq \frac92$.
 The $2N$ potential is AV18.}
\end{table*}

\begin{table*}[!] 
\begin{ruledtabular}
\begin{tabular}{l*{8}{c}}
& $0^+ \; (^1S_0)$  & $0^- \; (^3P_0)$ & $1^+ \; (^3S_1) $ &
  $1^- \; (^3P_1)$  & $1^- \; (^1P_1)$ & $1^- \; (\epsilon)$ & 
  $2^- \; (^3P_2)$  & $\sigma_t $
\\  \hline
$l_y, l_z \leq 0$ & -69.70 & & -63.50 & & & & & 0.950\\
$l_y, l_z \leq 1$ & 
-69.59 & 22.62 & -61.94 & 41.33 & 22.65 & -44.49 & 43.35 & 2.163 \\
$l_y, l_z \leq 2$ 
& -69.67 & 23.19 & -61.75 & 42.65 & 22.05 & -44.88 & 44.10 & 2.196  \\ 
$l_y, l_z \leq 3$
& -69.62 & 23.65 & -61.72 & 43.35 & 22.34 & -44.84 & 46.82 & 2.279 \\ 
$l_y, l_z \leq 4$ 
& -69.63 & 23.62 & -61.69 & 43.54 & 22.38 & -44.69  & 47.03 & 2.288 
\end{tabular}
\end{ruledtabular}
\caption{ \label{tab:2}
Same as in Table~\ref{tab:1} for increasing $l_y, l_z$ and
fixed $I \leq 4 + {}^3G_5$ and $J \leq \frac92$.
The restriction on $l_y$ is not applied to $3N$ partial waves with
total angular momentum and parity $J^\pi = \frac12^+$.}
\end{table*}

\begin{table*}[!] 
\begin{ruledtabular}
\begin{tabular}{l*{8}{c}}
& $0^+ \; (^1S_0)$  & $0^- \; (^3P_0)$ & $1^+ \; (^3S_1) $ &
  $1^- \; (^3P_1)$  & $1^- \; (^1P_1)$ & $1^- \; (\epsilon)$ & 
  $2^- \; (^3P_2)$  & $\sigma_t $
\\  \hline
$J \leq \frac12$ 
& -69.84 & 23.95 & -53.98 & 27.53 & 17.55 & -9.48  & 17.56 & 1.268   \\ 
$J \leq \frac32$ 
& -69.61 & 23.26 & -62.41 & 43.05 & 22.34 & -44.85  & 21.97 & 1.715  \\
$J \leq \frac52$ 
& -69.63 & 23.61 & -61.69 & 43.49 & 22.37 & -44.63  & 46.97 & 2.285  \\
$J \leq \frac72$ 
& -69.63 & 23.61 & -61.69 & 43.53 & 22.38 & -44.68  & 46.99 & 2.287  \\
$J \leq \frac92$ 
& -69.63 & 23.62 & -61.69 & 43.54 & 22.38 & -44.69  & 47.03 & 2.288
\end{tabular}
\end{ruledtabular}
\caption{ \label{tab:3} 
Same as in Table~\ref{tab:1} for increasing $J$ and fixed 
$I \leq 4 + {}^3G_5$  and $l_y, l_z \leq 4$.}
\end{table*} 

\begin{table*}[!] 
\begin{ruledtabular}
\begin{tabular}{l*{8}{c}}
& $0^+ \; (^1S_0)$  & $0^- \; (^3P_0)$ & $1^+ \; (^3S_1) $ &
  $1^- \; (^3P_1)$  & $1^- \; (^1P_1)$ & $1^- \; (\epsilon)$ & 
  $2^- \; (^3P_2)$  & $\sigma_t $
\\  \hline
AV18 & -66.12 & 20.75 & -58.48 & 40.09 & 20.73 & -44.50 & 42.51 & 2.331 \\ 
\quad Ref.~\cite{lazauskas:05a}
& -66.5 & 20.9 & -58.5 & 37.3 & 20.7 & -43.5 & 41.0  & 2.24 \\
\quad Ref.~\cite{lazauskas:05a}
& -66.3 & 20.6 & -58.7 & 38.6 & 20.5 & -45.5 &  40.1 & 2.24 \\ 
\quad rank 1 
        & -66.06 & 26.99 & -58.55 & 42.36 & 22.15 & -44.81 & 45.06 & 2.488 \\ 
CD Bonn & -64.63 & 18.97 & -57.40 & 39.44 & 20.20 & -44.94 & 42.47 & 2.283 \\ 
Nijmegen I 
        & -65.61 & 19.64 & -58.16 & 39.62 & 20.40 & -44.91 & 42.13 & 2.297 \\ 
Nijmegen II
        & -65.98 & 20.02 & -58.42 & 39.69 & 20.44 & -44.71 & 42.22 & 2.308 \\
INOY04  & -62.91 & 16.73 & -56.00 & 38.75 & 19.47 & -44.55 & 42.13 & 2.216 \\ 
N3LO    & -65.54 & 20.31 & -57.99 & 40.94 & 20.74 & -44.71 & 43.98 & 2.377 \\
\end{tabular}
\end{ruledtabular}
\caption{ \label{tab:4}
$\nH$ phase shifts, mixing parameter $\epsilon$, and total cross 
section $\sigma_t$ for AV18, CD-Bonn, Nijmegen I, Nijmegen II, INOY04,
 and N3LO potentials at  $E_n = 3.5$ MeV together with results from other 
calculations for AV18. We include $I \leq 4 + {}^3G_5$, $l_y, l_z \leq 4$,
and $J \leq \frac92$.}
\end{table*}

In Table~\ref{tab:1} we show $\nH$ phase shifts, $1^-$ mixing parameter
$\epsilon$, and total cross section $\sigma_t$ at $E_n = 4$ MeV neutron 
lab energy for increasing number of $2N$  partial waves. 
In all calculations we keep $ l_y,\,l_z \leq 4$ and $J \leq \frac92$. 
We apply additional restrictions that are different for $1+3$ and $2+2$
states. We include all $1+3$ states with $l_x + l_y \leq 8$ 
plus the states coupled to them by the tensor force;
the above restriction is not applied if $I \leq 2$.
We include all $2+2$ states with $l_x+l_y+l_z \leq 10$ plus states
coupled to them by the tensor force.
One finds that  at least $I \leq 3 + {}^3F_4$ 
is needed for a well converged calculation.
Likewise in Table~\ref{tab:2} we show similar results for
increasing  $ l_y,\,l_z$ keeping  $I \leq 4 + {}^3G_5$ and $J \leq \frac92$. 
At least  $ l_y,\, l_z \leq 3$ is needed to get quite satisfactorily 
converged results for the $P$-wave phase shifts, particularly ${}^3P_2$.
Finally in Table~\ref{tab:3} we show results for increasing $J$, keeping $
l_y,\, l_z \leq 4$ and $I \leq 4 +  {}^3G_5$. We find that the inclusion of
at least $J=\frac52$ \, $3N$ states is necessary without which $A_y$ has the
wrong sign. Compared with previous calculations the present work exceeds in the
number of $2N$, $3N$, and $4N$ partial waves included, 
providing very accurate results for all observables. 

In Table~\ref{tab:4} we show the results of the other calculations for 
AV18 at $E_n = 3.5$ MeV which were compiled in Ref.~\cite{lazauskas:05a}.
The present calculation confirms the work of the
Grenoble and Pisa groups (second and third lines, respectively) 
and clearly shows in the fourth line the shortcomings of
the rank one representation of realistic interactions calculated again using
the present numerical algorithm. As in the work of Ref.~\cite{fonseca:99a} 
the total cross section gets to be $\sigma_t = 2.49$ b which even slightly
overestimates the experimental value. 
Calculations with other potentials, i.e.,
charge-dependent (CD) Bonn \cite{machleidt:01a}, 
Nijmegen I and II \cite{stoks:94a},
inside-nonlocal outside-Yukawa (INOY04) potential by Doleschall
\cite{doleschall:04a,lazauskas:04a}, and 
the one  derived from chiral perturbation theory at 
next-to-next-to-next-to-leading order (N3LO) \cite{entem:03a},
show similar results for all phases although N3LO gives the largest $P$-wave
phases leading to $\sigma_t = 2.38$ b,
the closest to the experimental value at the
resonance peak using two-body interactions alone.

\begin{figure}[!]
\begin{center}
\includegraphics[scale=0.65]{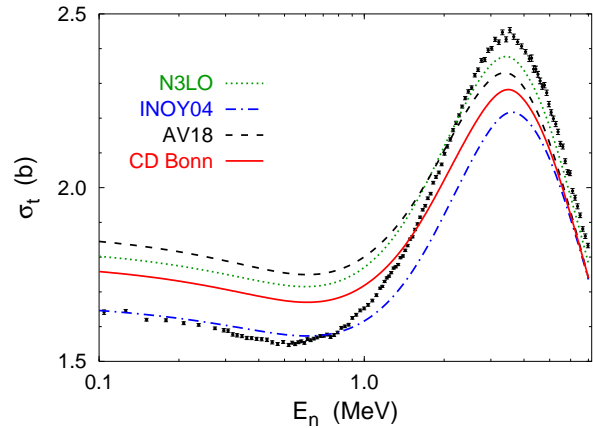}
\end{center} %\vspace{-5mm}
\caption{ \label{fig:nttot} (Color online)
Total cross section for $\nH$ scattering as function of neutron lab 
energy calculated with CD Bonn (solid curve), AV18 (dashed curve), 
INOY04 (dash-dotted curve), and N3LO (dotted curve) potentials.
Experimental data are from \Ref~\cite{phillips:80}.}
\end{figure}

In \Fig~\ref{fig:nttot} we show the total cross section  for $\nH$ 
scattering as a function of the neutron lab energy; 
for clarity we skip the Nijmegen I and II predictions since they
are between AV18 and CD Bonn. In the resonance
region all potentials fail to reproduce the experimental data though some do
better than others. As pointed out in Ref.~\cite{lazauskas:04a} the nonlocal
potential INOY04 that, by itself alone, leads to the correct triton 
binding energy and slightly overbinds the $\alpha$ particle, 
shows the lowest total cross section at the peak. On
the contrary CD Bonn and AV18 show higher total cross sections but also lower
triton and $\alpha$ particle binding energies. 

\begin{figure}[h]
\begin{center}
\includegraphics[scale=0.6]{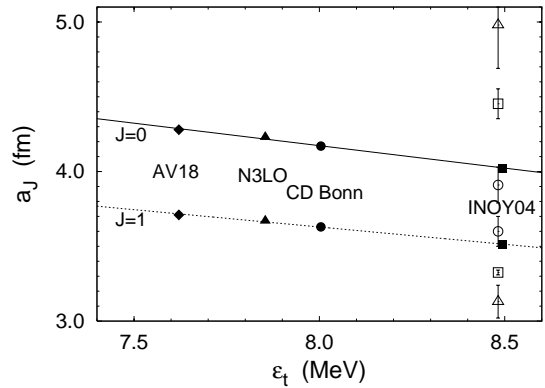}
\end{center} %\vspace{-5mm}
\caption{ \label{fig:e-a} Correlation between $\Hh$ binding energy
$\varepsilon_t$ and $\nH$ scattering lengths $a_0$ and $a_1$. 
The predictions for AV18 (diamonds), N3LO (triangles), 
CD Bonn (circles),  and INOY04 (squares) are shown.
The experimental data are from \Ref~\cite{seagrave:80} (open circles),
\Ref~\cite{rauch:85} (open triangles), and
\Ref~\cite{hale:90a} (open squares).}
\end{figure}

In Table~\ref{tab:5} we give the
values for the triton and $\alpha$ particle binding energies, 
singlet and triplet scattering lengths $a_0$ and $a_1$, and total cross 
section at $E_n = 0$ and 3.5 MeV. The results we get
for $a_0$ and $a_1$ agree with previous work for 
AV18~\cite{lazauskas:04a,lazauskas:05a}, 
and as shown in \Fig~\ref{fig:e-a} correlate with the triton binding energy.
Therefore interactions that lead to lower triton binding show 
the highest values for $a_0$ and $a_1$ and consequently the higher total 
cross sections at threshold. Nevertheless at $E_n = 3.5$ MeV this 
correlation gets destroyed as the behavior of N3LO shows.
Further studies are needed to understand the features of N3LO that give 
rise to this breaking of the correlation near the peak of the resonance.

\begin{table}[htbp] 
\begin{ruledtabular}
\begin{tabular}{l*{8}{c}}
 & $\varepsilon_t$ & $\varepsilon_{\alpha}$  & $a_0 $  &
$a_1 $  & $\sigma_t$ (0) & $\sigma_t$ (3.5)  \\  \hline   
AV18        & 7.621 & 24.24 & 4.28 & 3.71 & 1.88 & 2.33 \\
Nijmegen II & 7.653 & 24.50 & 4.27 & 3.71 & 1.87 & 2.31 \\
Nijmegen I  & 7.734 & 24.94 & 4.25 & 3.69 & 1.85 & 2.30 \\
N3LO        & 7.854 & 25.38 & 4.23 & 3.67 & 1.83 & 2.38 \\ 
CD Bonn     & 7.998 & 26.11 & 4.17 & 3.63 & 1.79 & 2.28 \\ 
INOY04      & 8.493 & 29.11 & 4.02 & 3.51 & 1.67 & 2.22 \\ 
\end{tabular}
\end{ruledtabular}
\caption{$\Hh$ and ${}^4\mathrm{He}$ binding energies 
$\varepsilon_t$ and $\varepsilon_{\alpha}$ (in MeV), 
$\nH$ scattering lengths $a_0$ and $a_1$ (in fm), 
and $\nH$ total cross section $\sigma_t$ (in barns) at $E_n = 0$ and 3.5 MeV
neutron lab energy for different $2N$ potentials.}
\label{tab:5}
\end{table}

\begin{figure*}[!]
\begin{center}
\includegraphics[scale=0.6]{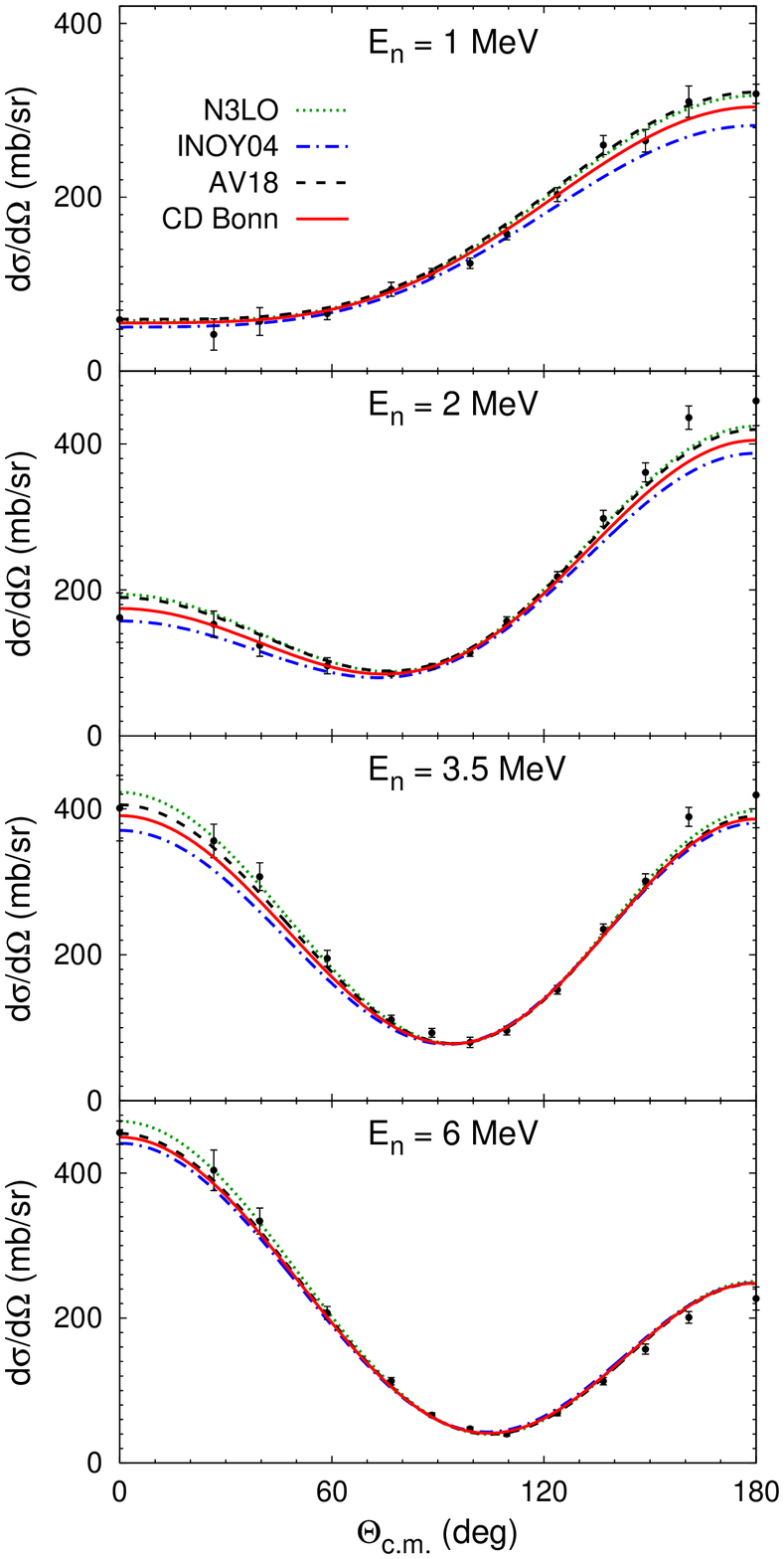} \quad
\includegraphics[scale=0.6]{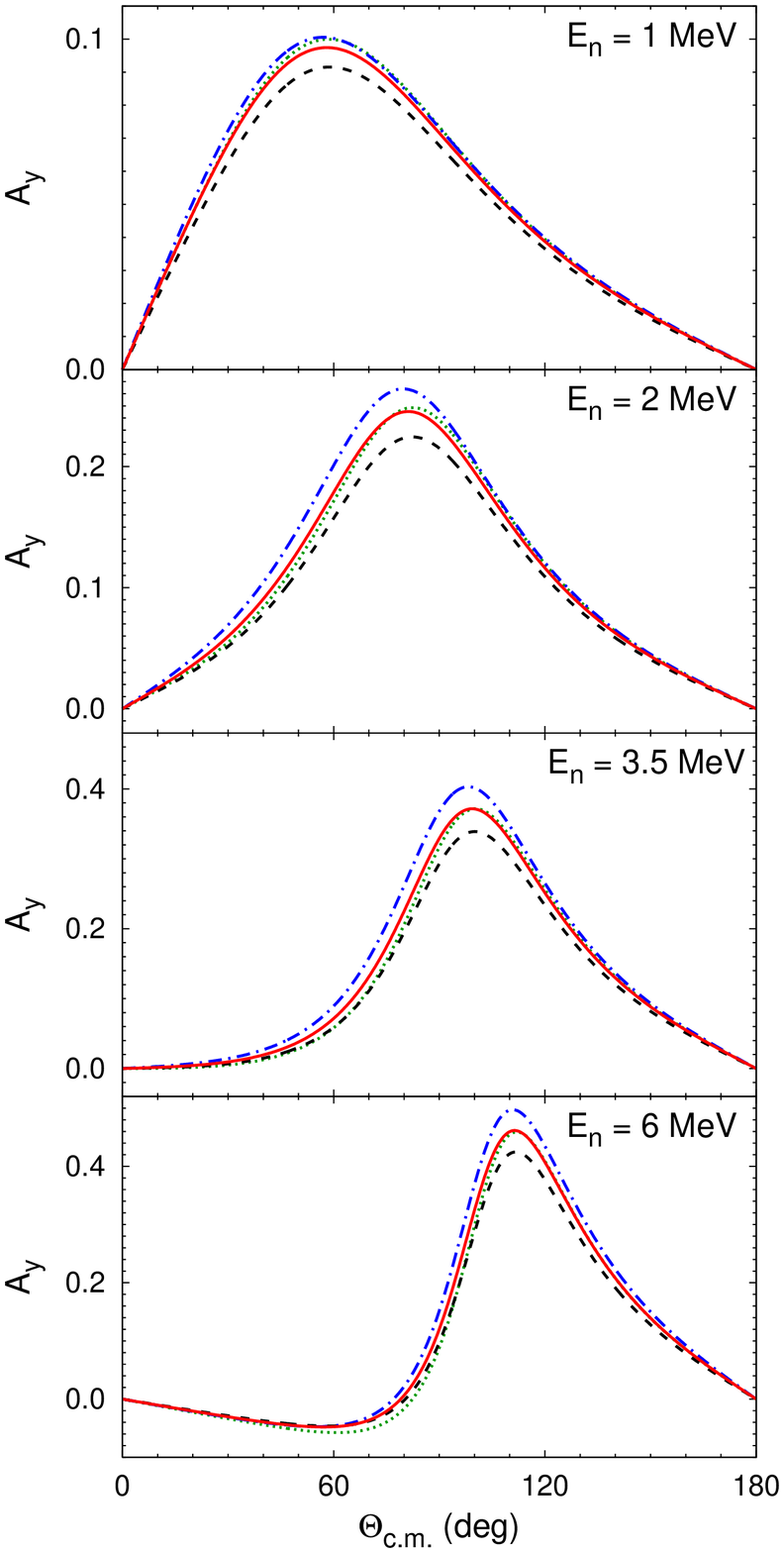}
\end{center} %\vspace{-5mm}
\caption{ \label{fig:dcs} (Color online) Differential cross section
and neutron analyzing power for $\nH$ scattering at $E_n = 1$, 2, 3.5, 
and 6 MeV neutron lab energies as functions of c.m. scattering angle.
Curves as in Fig.~\ref{fig:nttot}.
Experimental data are from \Ref~\cite{seagrave:60}.}
\end{figure*}

In  Fig.~\ref{fig:dcs} we show the differential cross section 
$d\sigma/d\Omega$ and the neutron analyzing power $A_y$ for
$\nH$ scattering at neutron lab energies of 1, 2, 3.5, and 6 MeV.
In order to get fully converged results we take into account all $\nH$
channel states with  orbital angular momentum $L \leq 3$.
The predictions of the four potentials differ mostly at forward and
backward angles for the differential cross section and
around the peak for the analyzing power. 
It is not obvious to us that the disagreement with the total cross section
data shown in Fig.~\ref{fig:nttot} is compatible with the discrepancies
we observe relative to the differential cross section data.
Therefore it would be recommended that some of the experiments be 
repeated at specific energies and $A_y$ measured in order to further
understand the implications of the $2N$ force models.

One important observation that comes out of these calculations is 
the increased sensitivity of $4N$ observables to changes in the $2N$ 
interaction. The variations due to the $2N$ potential
 at the maximum of $\nH$ $A_y$ lead to about 16\% fluctuations
which are larger than the 10\% fluctuations observed at the peak of $A_y$
in low energy $\nd$ scattering.
This indicates that the $4N$ system is more sensitive to off-shell 
differences of the $2N$ force than the $3N$ system.

Finally, in Table \ref{tab:Pmod} we investigate the possible correlations
between the $A_y$-puzzle  in low energy $\nd$ scattering
and the underestimation of $\sigma_t$ in $\nH$ scattering in the 
resonance region. The experimental data for $A_y$ in $\nd$ scattering 
can be accounted for by a calculation with modified interactions
in $2N$ ${}^3P_I$ waves \cite{tornow:98a,doleschall:04a}.
We use two models. The first one, AV18', is taken from Ref.~\cite{tornow:98a};
it corresponds to the AV18 potential
that in ${}^3P_I$ waves is multiplied by strength factors
0.96, 0.98, and 1.06 for $I=0$, 1, and 2, respectively.
The second one, INOY04', is taken from Ref.~\cite{doleschall:04a}
and differs from INOY04 by ${}^3P_I$ wave parameters.
Although both modified potentials provide quite satisfactory
description of vector analyzing powers in low energy $\nd$ scattering,
they are incompatible with present day $2N$ data basis. 
E.g., the ${\chi}^2 / \mathrm{datum}$ values with respect to the $pp$ data, 
estimated using the Nijmegen error matrix \cite{stoks:93a},
i.e., by comparing to the Nijmegen phase shifts rather than to data directly,
are 3.5 for INOY04' and 4.4 for AV18' potentials.
However, those modifications of the potentials are unable to resolve
the  $\sigma_t$ discrepancy in $\nH$ scattering. The $\sigma_t$ is slightly
increased for AV18' but it gets even lower for INOY04', indicating
that $\sigma_t$ depends on the $2N$ ${}^3P_I$ wave interaction in a different
way than the $A_y$ in the $\nd$ scattering.

\begin{table*}[!] 
\begin{ruledtabular}
\begin{tabular}{l*{8}{c}}
& $0^+ \; (^1S_0)$  & $0^- \; (^3P_0)$ & $1^+ \; (^3S_1) $ &
  $1^- \; (^3P_1)$  & $1^- \; (^1P_1)$ & $1^- \; (\epsilon)$ & 
  $2^- \; (^3P_2)$  & $\sigma_t $
\\  \hline
AV18 & -66.12 & 20.75 & -58.48 & 40.09 & 20.73 & -44.50 & 42.51 & 2.331 \\ 
AV18'& -66.06 & 20.46 & -58.39 & 40.50 & 20.86 & -44.70 & 43.82 & 2.375 \\
INOY04 & -62.91 & 16.73 & -56.00 & 38.75 & 19.47 & -44.55 & 42.13 & 2.216 \\ 
INOY04'& -63.04 & 15.67 & -56.16 & 37.41 & 19.06 & -43.06 & 42.21 & 2.191 
\end{tabular}
\end{ruledtabular}
\caption{ \label{tab:Pmod}
$\nH$ phase shifts, $1^-$ mixing parameter $\epsilon$, and total cross 
section $\sigma_t$ at  $E_n = 3.5$ MeV
for original AV18 and INOY04 potentials and their
versions AV18' and INOY04' with modified $P$-waves.}
\end{table*}

%\clearpage
\section{Conclusions \label{sec:conc}}

In the present paper we developed a new numerical approach to solve
four-nucleon scattering equations in momentum-space. 
The method uses no uncontrolled approximations, is numerically
very efficient and therefore can include very large number of
partial waves, thereby yielding well converged and very precise results.
The developed approach is applied to $\nH$ scattering below 
three-body breakup threshold. 
The calculations with various realistic $2N$ potentials underestimate
the total $\nH$ cross section data in the resonance region 
as already found by other groups. However, probably due to the inclusion
of more partial waves, the numbers we get are slightly higher
and closer to the data; they also depend on the choice
of the  $2N$ potential.
The new results also show that $4N$ observables are more sensitive
than $3N$ observables  to the off-shell nature of the $2N$ interaction.
Furthermore, the modifications that are required to introduce at the level
of the ${}^3P_I$ $2N$ partial waves to remove the discrepancies in $\nd$
$A_y$ at low energy, do not remove the disagreement observed in
the total $\nH$ cross section around $E_n = 3.5$ MeV.
Finally, to understand the compatibility between existing $\nH$ total and
differential cross section  data it would be advisable to repeat some
of those experiments at specific energies.

\begin{acknowledgments}
The authors thank R.~Lazauskas for valuable discussions and 
for providing benchmark results.
A.D. is supported by the Funda\c{c}\~{a}o para a Ci\^{e}ncia e a Tecnologia
(FCT) grant SFRH/BPD/14801/2003
and A.C.F. in part by the FCT grant POCTI/ISFL/2/275.
\end{acknowledgments}

%%%%%%%%%%%%%%%%%%%%%%%%%%%%%%%%%%%%%%%%%%%%%%%%%%%%%%%%%%%%%%%%%%%%%%%%%%%%%

\begin{appendix}
\section{} 

As mentioned in Section~\ref{sec:res} we present here our results for the
binding energy of ${}^4\mathrm{He}$ and $\nH$ phase shifts obtained with 
Mafliet-Tjon potential as well as the numerical stability check of our results.
In Table~\ref{tab:6} we show the $\alpha$ particle
binding energy for increasing number
of $2N$ partial waves and compare with previous works. Results with
AV8' are calculated without the Coulomb interaction in order to
compare with Ref.~\cite{kamada:01a}. On the other hand calculations from
Ref.~\cite{nogga:02a,viviani:06a} with CD Bonn include coupling between total
isospin $T=0,\,1$ and 2 states while we consider only $T=0$.
In contrast to our scattering calculations, here we perform the isospin
averaging not for the t-matrix but for the potential 
like it has been done in calculations of Ref.~\cite{lazauskas:04a}.
Overall these results indicate that our algorithm is accurate and reliable.
 
\begin{table*}[!] 
\begin{ruledtabular}
\begin{tabular}{l*{7}{c}}
 & $I \leq$ 1  & $I \leq$ 2 & $I \leq$ 3  & $I \leq$ 4 
& $I \leq$ 5 & $I \leq$ 6 & 
Other work: Refs.~\cite{nogga:02a,kamada:01a,lazauskas:04a,viviani:06a}
\\ \hline
AV8'    & 23.08 & 25.16 & 25.69 & 25.85 & 25.90 & 25.91 & 25.90 - 25.94 \\ 
AV18    & 22.30 & 23.75 & 24.15 & 24.20 & 24.23	& 24.24 & 24.22 - 24.25  \\
CD Bonn & 25.03 & 25.95 & 26.07 & 26.10 & 26.11 & 26.11 & 26.13 - 26.16 \\
INOY04  & 28.68 & 29.09 & 29.10 & 29.11 & 29.11 & 29.11 & 29.11 
\end{tabular}
\end{ruledtabular}
\caption{${}^4\mathrm{He}$ Binding Energy (MeV) for increasing number of
$2N$ partial waves characterized by maximal total angular momentum $I$.}
\label{tab:6}
\end{table*} 

\begin{table*}[!]
\begin{ruledtabular}
\begin{tabular}{l*{6}{c}}
 & $L=0$, $S=0$ & $L=1$, $S=0$ & $L=2$, $S=0$ & 
$L=0$, $S=1$ &$L=1$, $S=1$ & $L=2$, $S=1$ \\ \hline
$E_n = 2.0$ MeV  & 50.93 & 17.19 & -0.37 &  -45.65 & 22.56 & -0.57 \\
\quad Ref.~\cite{lazauskas:phd} 
& 51.1 & 17.2 & -0.37 & -45.8 & 22.6 & -0.58 \\
$E_n = 3.5$ MeV & -64.53 & 28.00 & -1.39 & -58.17 & 40.51 & -0.94  \\
\quad Ref.~\cite{lazauskas:phd} 
& -64.6 & 28.0 & -1.40 & -58.2 & 40.5 & -0.89 \\
$E_n = 5.0$  MeV & -74.33 & 34.06 & -2.17 & -67.30 & 50.56 & -1.53 \\
\quad Ref.~\cite{lazauskas:phd} 
& -74.4 & 34.0 & -2.24 & -67.4 & 50.5 & -1.59 
\end{tabular}
\end{ruledtabular}
\caption{$\nH$ phase shifts at
different neutron lab energies for the Mafliet-Tjon potential.}
\label{tab:7}
\end{table*}

For $\nH$ scattering we compare in Table~\ref{tab:7} the results of our
calculations with those of the Grenoble group~\cite{lazauskas:phd}. 
Again our phase shifts agree within a few tenth of a degree or better 
leading to identical total cross sections over a wide range of energies. 

\begin{table*}[!] 
\begin{ruledtabular}
\begin{tabular}{l*{8}{c}} $N_k$ 
& $0^+ \; (^1S_0)$  & $0^- \; (^3P_0)$ & $1^+ \; (^3S_1) $ &
  $1^- \; (^3P_1)$  & $1^- \; (^1P_1)$ & $1^- \; (\epsilon)$ & 
  $2^- \; (^3P_2)$  & $\sigma_t $
\\  \hline 
14 & -70.03 & 23.78 & -62.01 & 43.49 & 22.35 & -44.58 & 46.93 & 2.290 \\
15 & -69.57 & 23.57 & -61.61 & 43.64 & 22.40 & -44.72 & 47.18 & 2.292 \\
16 & -69.43 & 23.51 & -61.51 & 43.63 & 22.38 & -44.75 & 47.19 & 2.290 \\
18 & -69.72 & 23.66 & -61.76 & 43.55 & 22.39 & -44.67 & 47.03 & 2.289 \\
20 & -69.63 & 23.62 & -61.69 & 43.54 & 22.38 & -44.69 & 47.03 & 2.288 \\
22 & -69.68 & 23.64 & -61.73 & 43.53 & 22.39 & -44.68 & 47.01 & 2.288 \\
24 & -69.66 & 23.64 & -61.73 & 43.53 & 22.39 & -44.68 & 47.00 & 2.288 \\
25 & -69.67 & 23.64 & -61.73 & 43.53 & 22.39 & -44.68 & 47.00 & 2.288
\end{tabular}
\end{ruledtabular}
\caption{$\nH$ phase shifts, $1^-$ mixing parameter $\epsilon$,
and  the total cross section $\sigma_t$ at $E_n = 4$ MeV  
for increasing number of  meshpoints $N_k$ for the momenta $k_x$, $k_y$,
and $k_z$. The $2N$ potential is AV18.}
\label{tab:Nk} 
\end{table*}

In Table~\ref{tab:Nk} we demonstrate the stability of our results 
increasing the number $N_k$ of momentum meshpoints. All calculations
use AV18 potential with $2N$ partial waves $I \leq 4 + {}^3G_5 $,
  $l_y,\,l_z \leq 4$  and $J \leq \frac92$.
The results with  $N_k = 20$ which is the choice for calculations
of Tables \ref{tab:1} --- \ref{tab:3}
are converged to better than 0.05 deg for all phase shifts.

\end{appendix}
%\clearpage
%%%%%%%%%%%%%%%%%%%%%%%%%%%%%%%%%%%%%%%%%%%%%%%%%%%%%%%%%%%%%%%%%%%%%%%%%%%%%
\bibliographystyle{prsty}
%\bibliography{abbrev,pre80,80-89,90-99,200x,clmb,ad,4N,hann,book,numerics}
%\end{document}

\end{document}